\documentclass{aa} 

\usepackage{graphicx}
\usepackage{txfonts}
\begin{document}

   \title{Vacuum-UV photodesorption from compact Amorphous Solid Water : photon energy, isotopic and temperature effects}

  \author{J.-H. Fillion \inst{1}, R. Dupuy  \inst{1}, G. F\'eraud  \inst{1}, C. Romanzin  \inst{2}, L. Philippe \inst{1}, T. Putaud  \inst{1},
  V. Baglin  \inst{3}, R. Cimino  \inst{4}, P. Marie-Jeanne  \inst{1}, P. Jeseck \inst{1}, X. Michaut \inst{1}  
           \and
          M. Bertin\inst{1}{}
          }

   \institute{Sorbonne Universit\'e, Observatoire de Paris, Universit\'e PSL, CNRS, LERMA F-75005 Paris\\
              \email{jean-hugues.fillion@sorbonne-universite.fr}
         \and
             Institut de Chimie Physique, UMR8000 CNRS, Université Paris-Saclay, 91405, Orsay, France
         \and CERN, CH-1211 Geneva 23, Switzerland
         \and Laboratori Nazionali di Frascati (LNF)-INFN I-00044 Frascati, Italy
           }

   \date{Received XXXX, 2019; accepted XXXX, XXXX}

 
  \abstract
   {Vacuum-UV (VUV) photodesorption from water-rich ice mantles coating interstellar grains is known to play an important role on the gas-to-ice ratio in stars and planets formation regions. Quantitative photodesorption yields from water ice are crucial for astrochemical models.}
   {The aims is to provide the first quantitative photon-energy dependent photodesorption yields from water ices in the VUV. This information is important for the understanding of the photodesorption mechanisms and to account for the variation of the yields under interstellar irradiation conditions.}
   {Experiments have been performed on the DESIRS beamline at the SOLEIL synchrotron facility (St Aubin, France), delivering tunable VUV radiation, using the ultra-high SPICES (Surface Processes and ICES) vacuum chamber. Thick compact amorphous solid water ice (H$_2$O and D$_2$O) grown onto a cold Au substrate have been irradiated from 7 to 13.5 eV.  Quantitative yields have been obtained by detection into the gas phase with mass-spectrometry for sample temperatures ranging from 15 K to 100 K.}
   {Photodesorption spectra  of H$_2$O (D$_2$O), OH (OD), H$_2$ (D$_2$) and O$_2$ peak around 9-10 eV and decrease at higher energies. Average photodesorption yields of intact water at 15 K are 5 $\times$ 10$^{-4}$ molecule/photon for H$_2$O and 5 $\times$ 10$^{-5}$  molecule/photon for D$_2$O over the 7-13.5 eV range.  The strong isotopic effect can be explained by a differential chemical recombination between OH (OD) and H (D) photofragments originating from lower kinetic energy available for the OH photofragments upon direct water photodissociation and/or possibly by an electronic relaxation process. It is expected to contribute to water fractionation during the building-up of the ice grain mantles in molecular clouds and to favor OH-poor chemical environment in comet-formation regions of protoplanetary disks. The yields of all the detected species except OH (OD) are enhanced above (70 $\pm$10) K, suggesting an ice restructuration at this threshold temperature.}
   {}

   \authorrunning{Fillion et al.}
   \titlerunning {Photodesorption from c-ASW}
   \maketitle
%

\section{Introduction}

  The photodesorption from water ice by Vacuum-UltraViolet (VUV) photons ($\lambda$ < 200 nm) is known to play an important role in variety 
  of astrophysical cold environments ranging from  star and planets formation regions of the Interstellar Medium (ISM)
  to icy bodies of the solar system. The desorption induced by VUV photons is a non-thermal process, allowing atoms and molecules to be ejected into the gas phase from low temperature ices, especially below their condensation temperature. In the ISM, this phenomenon can deeply affect  the gas-to-ice ratio in cold environments.   Astrophysical models have shown that photodesorption from external stellar VUV radiation (6-13.6 eV) 
  can prevent the formation of the water ice mantle at the surface of dust grains located at the edge of molecular clouds and supply most of the gas phase water abundance at intermediate depths into the cloud \citep{hollenbach_water_2009, furuya_water_2015}. 
  Deeper inside molecular clouds, the secondary VUV radiation induced by cosmic rays, although  $10^3$ times less intense than the interstellar standard radiation field governing the photochemistry at the edge of molecular clouds, is a non-thermal desorption agent \citep{caselli_first_2012}, along with the desorption induced by exothermic chemistry from grains at the very early stages of the cloud formation  \citep{cazaux_dust_2016}. The desorption directly induced by cosmic ray impact onto icy grain mantles is also strongly contributing to these non-thermal phenomena \citep{dartois_heavy_2015, dartois_cosmic_2018}.  Non-thermal desorption are believed to play an important role in  protoplanetary environments by maintaining molecules into the gas phase in the surface or intermediate regions of the disk \citep{Aikawa_molecular_1999, willacy_importance_2000}.  \citet{Dupuy_x-ray_2018} recently highlighted the efficient role played by soft X-ray photons to induce the non-thermal desorption of neutral water from water ice.  In the comet formation region of the disk, VUV photodesorption from cold grains is supposed to be a source of reactive species, such as OH, which alter drastically the gas phase chemical composition \citep{chaparro_molano_role_2012}. Knowing the nature and photodesorption yields of the desorbing photofragments is therefore of primary importance. Moreover, VUV photodesorption of water at the disk surface, followed by recondensation to lower heights in the disk has been proposed to be at the origin of the amorphous structure of the ice in the outer solar nebula \citep{ciesla_phases_2014}.  Strong VUV radiation may also affect deeply the water ice distribution by pushing the snow line farther into the disk midplane \citep{terada_multi-epoch_2017}.  In the outer part of protoplanetary disks, chemical models show that most of the water is released into the gas phase through VUV photodesorption and can explain an enhancement of water vapor \citep{walsh_chemical_2010}.  In the disk of the young star TW hydrae, photodesorption is considered as the dominant form of desorption in cold unshielded regions where thermal desorption can be totally neglected \citep{fogel_chemistry_2011, salinas_first_2016}. Consequently, water lines emission from this object are sensitive to the absolute photodesorption yields employed in the disk model \citep{kamp_uncertainties_2013}.  Finally, photodesorption of water has also been considered to explain the ortho-to-para ratio (OPR) observed from water emission lines around TW Hydrae \citep{hogerheijde_detection_2011,salinas_first_2016} or in the Orion photon dominated region \citep{putaud_water_2019}.

  In the laboratory, numerous water photodesorption experiments have been performed since the pioneering studies of \citet{westley_photodesorption_1995, westley_ultraviolet_1995}.  \citet{oberg_photodesorption_2009} have studied the dependence of a number of physical parameters on the H$_2$O and D$_2$O photodesorption yields, such as ice thickness, ice temperature and morphology, VUV photon flux and fluence.  In these experiments,  absolute yields were derived from the water ice depletion measured in the solid phase, using reflection-absorption IR spectroscopy.
  \citet{cruz-diaz_new_2018} have reported new measurements of UV-photodesorption from H$_2$O and D$_2$O ices using a calibrated quadrupole mass
 spectrometer in order to detect directly, into the gas phase, intact water together with photofragments and other species formed by surface photochemistry.
 In all the above mentioned experiments, VUV photons were generated using a low pressure microwave-powered plasma source of H$_2$,
 producing an emission spectrum peaking at atomic H-Lyman-$\alpha$ (10.2 eV) superimposed on a continuum of molecular
 emissions (6 -11 eV) dominated by a large Lyman band of H$_2$  (B $^1\Sigma_u$ - W $^1\Sigma_g$) centered at 7.8 eV \citep{es-sebbar_optimization_2015, munoz_caro_uv-photoprocessing_2003}.
 Beside the above broadband excitation studies, the photodesorption from water ice films has also been studied at various selected energies using monochromatic sources.
 Experiments most probably involving multiphoton excitation processes have been carried-out using intense nanosecond laser pulses in the visible UV-range \citep{Nishi_photodetachment_1984, bergeld_photo_2006}. Direct excitation in the VUV at 9.8 eV (126 nm) and 7.2 eV (172 nm) have also been explored using Ar and Xe excimer lamps \citep{watanabe_measurements_2000}.
 More recently, the investigation of water photodesorption in the VUV using excimer lasers at 6.4 eV (193 nm) and 7.9 eV  (157 nm) coupled to a direct probe of the ejected species by
 Resonance Enhanced Multiphoton Ionisation (REMPI) have provided much insights on the desorption mechanisms from amorphous solid water at 100 K. \citet{yabushita_photochemical_2013} have reviewed these experimental results. Similar experiments, with slightly different conclusions on the kinetic energy released, have also been performed on thick water ice \citep{desimone_mechanisms_2013, desimone_o_2014}
 and compared to photodesorption from thinner Amorphous Solid Water (ASW) layers deposited on lunar substrates \citep{desimone_mechanisms_2013, desimone_h2o_2015}. All the above results could be faced to a valuable series of results from molecular dynamics simulations studying the desorption events  from ASW at low (10 K) and high temperature (90-100 K), which originate from initial photodissociation in the 7-9 eV energy range  \citep {andersson_photodesorption_2008, arasa_molecular_2010, andersson_theoretical_2011}. Theses studies include desorption from samples with different morphologies (amorphous vs crystalline) \citep{andersson_photodissociation_2005, andersson_molecular-dynamics_2006, crouse_photoexcitation_2015} 
 and isotopic effects studies \citep{arasa_molecular_2011, koning_isotope_2013,  arasa_photodesorption_2015}.
 
 The various excitation energies used in all these studies make direct quantitative comparison of the desorption yields difficult, because the way different excited states behave with respect to desorption is largely unknown. It has been shown, in the case of weakly bound species,  that
 photodesorption yields can indeed be strongly energy-dependent \citep{fayolle_co_2011, fayolle_wavelength-dependent_2013, bertin_indirect_2013, fillion_wavelength_2014, dupuy_efficient_2017}.
  Hydrogen discharge lamps are well-suited to simulate  the secondary VUV emission induced by cosmic rays inside dense molecular clouds and provide good estimate of the average desorption yields. However and despite the numerous experimental and theoretical studies mentioned above, no investigation of the desorption yields and mechanisms in the 10-13.5 eV range have been reported, though these energies could contribute to the desorption at the edges of molecular clouds or in protoplanetary disks. In order to model a specific region of space, one should ideally use monochromatic photodesorption
 yields weighted by the prevailing local VUV fields, but there is a lack of information concerning the photon-energy dependence of the photodesorption yields from water ice.\\

In this study, the photodesorption from compact ASW grown at 100 K has been investigated using synchrotron radiation enabling VUV irradiation from 7 to 13.5 eV for sample temperatures ranging from 15 K to 100 K. Water and desorbed photofragments are detected directly into the gas phase
 by mass spectrometry. The behavior of H$_2$O and D$_2$O pure ice samples under irradiation are systematically compared at various sample temperatures ranging from 15 K to 100 K, with the aim  to provide insights on the desorption mechanisms. Several possible desorption mechanisms are discussed to explain the strong isotopic effect observed. The absolute desorption yields  are compared to previous studies and astrophysical implications are discussed in the last section.

\section{Methods}

Experiments are performed in the SPICES (Surface Processes \& ICES) set-up developed at Sorbonne Université and Observatoire de Paris (France). This apparatus has been described in detail previously \citep{doronin_adsorption_2015}. It consists of an Ultra-High Vacuum (UHV) analysis chamber ($10^{-10}$ mBar) containing a polycrystalline gold surface mounted on a rotatable cold head that can be cooled down to 15 K using a closed cycle  helium cryostat. Ices of H$_2$O and D$_2$O (high purity, liquid chromatography standard from Fluka) are prepared by exposing the cold surface at 100 K to a partial pressure of gas using a tube positioned a few millimeters away from the surface, allowing growth of about 20 ML of compact Amorphous Solid Water (c-ASW) without increasing the chamber pressure to more than a few 10$^{-9}$ mBar \citep{fillion_d2_2009}. When D$_2$O is used, the experiment is flushed with 5$\times$10$^{-8}$ mbar of D$_2$O vapour during 20 minutes in order to saturate the walls with deuterium and thus avoid isotopic exchanges. The D/H purity is finally checked using Reflection Absorption Infrared Spectroscopy (RAIRS), with which O-D and O-H stretching bands of the deposited ices can be distinguished. Once the sample is condensed, the tube is moved away from the surface and the surface is rotated back towards the detector and the photon beam.
 Sample temperature can then be varied from 15 K to 100 K to probe the temperature effect on the photodesorption yields, while the ice is expected to keep its compact structure.  Ice thicknesses are controlled with a precision better than 1 monolayer (ML) via a calibration using the temperature programmed desorption (TPD) technique, as detailed in \citet{doronin_adsorption_2015}. The release of species into the gas phase is monitored by means of a Quadrupole Mass  Spectrometer (QMS, model QMS200, Balzers).

The SPICES set-up has been coupled to the undulator-based DESIRS (Dichroïsme Et Spectroscopie par Interaction avec le Rayonnement Synchrotron) beamline \citep{nahon_desirs:_2012} at the SOLEIL synchrotron facility (St Aubin, France) which provides monochromatic, tunable VUV light for irradiation of our ice samples. Higher harmonics of the undulator are suppressed using a Krypton gas filter, after which the beam is directly reflected toward the ice sample. The sample is placed out of the focus plane of the beamline spot in order to get an irradiated surface area of about 1 cm$^2$ on the sample. 
The window-free coupling of the set-up to the beamline enables investigations up to 13.6 eV, a well-known cut-off of interstellar VUV photons due to the ionization of Hydrogen. This setting yields very stable harmonic-free photon fluxes of 10$^{14-15}$  photons.cm$^{-2}$.s$^{-1}$ in a 1 eV bandwidth  \citep{nahon_desirs:_2012}. 

To acquire Photon-Stimulated Desorption (PSD) spectra, the relative amount of photodesorbed molecules are recorded by the QMS at each selected energy from 7 to 13.5 eV by steps of 0.5 or 1 eV. Each photon energy irradiation lasts about 20 s (shutter ON) which is sufficiently longer than the dwell time of the QMS (0.5 s). Each irradiation period is followed by 30 s acquisition without irradiation (shutter OFF)  in order to measure the
background level on each mass. The photon energies are successively scanned using an arbitrary order, with the aim to avoid any bias linked to the evolution of the sample composition with photon fluence. An average value is extracted when the same photon energy is probed several times. Photon fluxes are measured using a Si photodiode (SXUV-100, IRD), and varies from 1.5 $\times$ 10$^{15}$ photons.s$^{-1}$.cm$^{-2}$ at 7 eV (where the absorption cross sections are extremely small) to 2 $\times$ 10$^{14}$ photons.s$^{-1}$.cm$^{-2}$ above 10 eV (where the ice absorption cross sections increase significantly). Under these conditions, the average total photon fluence received by the sample at the end of a typical scan is $\lesssim$ 6 $\times$10$^{16}$ photons.cm$^{-2}$. PSD were performed for different substrate temperatures ranging from 15 K to 100 K. 

Once corrected from the background contribution and divided by the photon flux, the QMS signal I$_i$ of a species $i$ is converted to absolute photodesorption yields at each photon energy.
Absolute yields are calibrated against the well-known CO photodesorption yields \citep{fayolle_co_2011}. Thus, in the case of H$_2$O, we get:
 $$Y(H_2O) = f_{CO} \times \frac{\sigma(H_2O^+/H_2O)}{\sigma(CO^+/CO)} \times \frac{F(H_2O)}{F(CO)} \times I_{H_2O} $$
 
 where $f_{CO}$ is the calibration factor for CO obtained with the same experiment and the same relative position of the different elements (sample, photon beam, QMS), $\sigma(i^+/i)$ is the partial electron-impact ionization cross-section of the ion relatively to its parent molecule, $F(i)$ is the apparatus function of our QMS for species $i$ ($F(CO)$ is set to 1). These two correction factors take into account the molecular dependence of the efficiency by the ionization chamber in the QMS head, and the intrinsic apparatus function of our QMS.
For others species, the signal at a given m/z ratio is corrected, when relevant, from the contribution of fragments ions originating from
 dissociative electron-impact ionization into the QMS head of heavier species. For example, OH signal is corrected from the cracking of the photodesorbed H$_2$O into OH$^+$ in the ion source of the mass spectrometer. This correction is made by considering the photodesorbed H$_2$O signal, from which the OH$^+$ cracking signal can be estimated using the mass filter apparatus function and the dissociative electron impact ionization cross section of OH$^+$ from  H$_2$O given by \citet{Straub_absolute_1998}.
 
The absolute photodesorption yields uncertainties originate from several sources. One source of uncertainty is linked to the amplitudes of noise of the (photo)desorption and background signals respectively, that is the signal-to-noise ratio. This uncertainty is taken into account and characterized by the error bars plotted in the figures, giving that way the range of the photodesorption yields measured with equal probability. A second source of uncertainty originates from the calibration procedure. This includes the uncertainties associated to the ionisation cross sections, that of the apparatus function of the mass spectrometer, and the uncertainty associated with the calibration of the incident photon flux. We estimate that overall, the uncertainty on the absolute values related to the calibration procedure amounts to $\pm$50\%. This illustrates how difficult is the determination of absolute photodesorption yields from the experiment. Finally, a third source of uncertainties can be associated to the possible structural and chemical evolutions of the sample during irradiation, since in this experiment the sample can’t be renewed between each data point. This last source of uncertainty has been estimated for each desorbing species and will be discussed in the result section.
\section{Results}

\begin{figure}
     \centering
   \includegraphics [width=8.5cm]{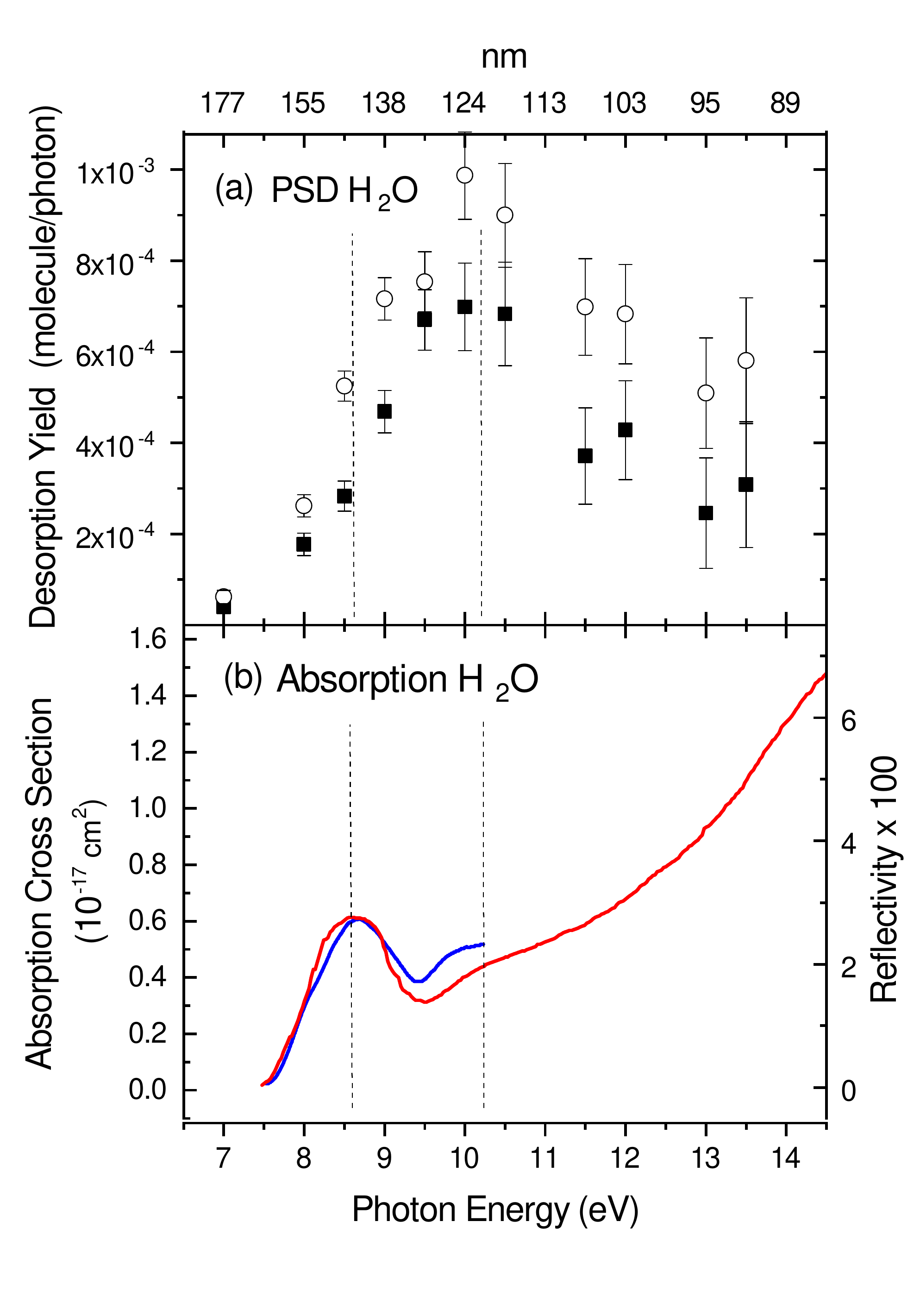}
   \caption{(a) Photodesorption yields (molecule per incident photon) of intact H$_2$O at 15 K (black squares) and at 100 K (open circles) from 20 ML of c-ASW films (b)  VUV-absorption cross sections of pure water ice deposited at 8 K from \citet{cruz-diaz_vacuum-uv_2014-1}
    (blue curve) and reflectivity spectrum of amorphous ice at 80 K from \citet{kobayashi_optical_1983}  (red curve). The reflectivity spectrum has been rescaled against the  absorption spectrum at 8.3 eV. The dotted vertical lines point the position of the first two excitonic bands.}
        \label{figH2O}
    \end{figure}

    \begin{figure}
     \centering
   \includegraphics [width=8.5cm]{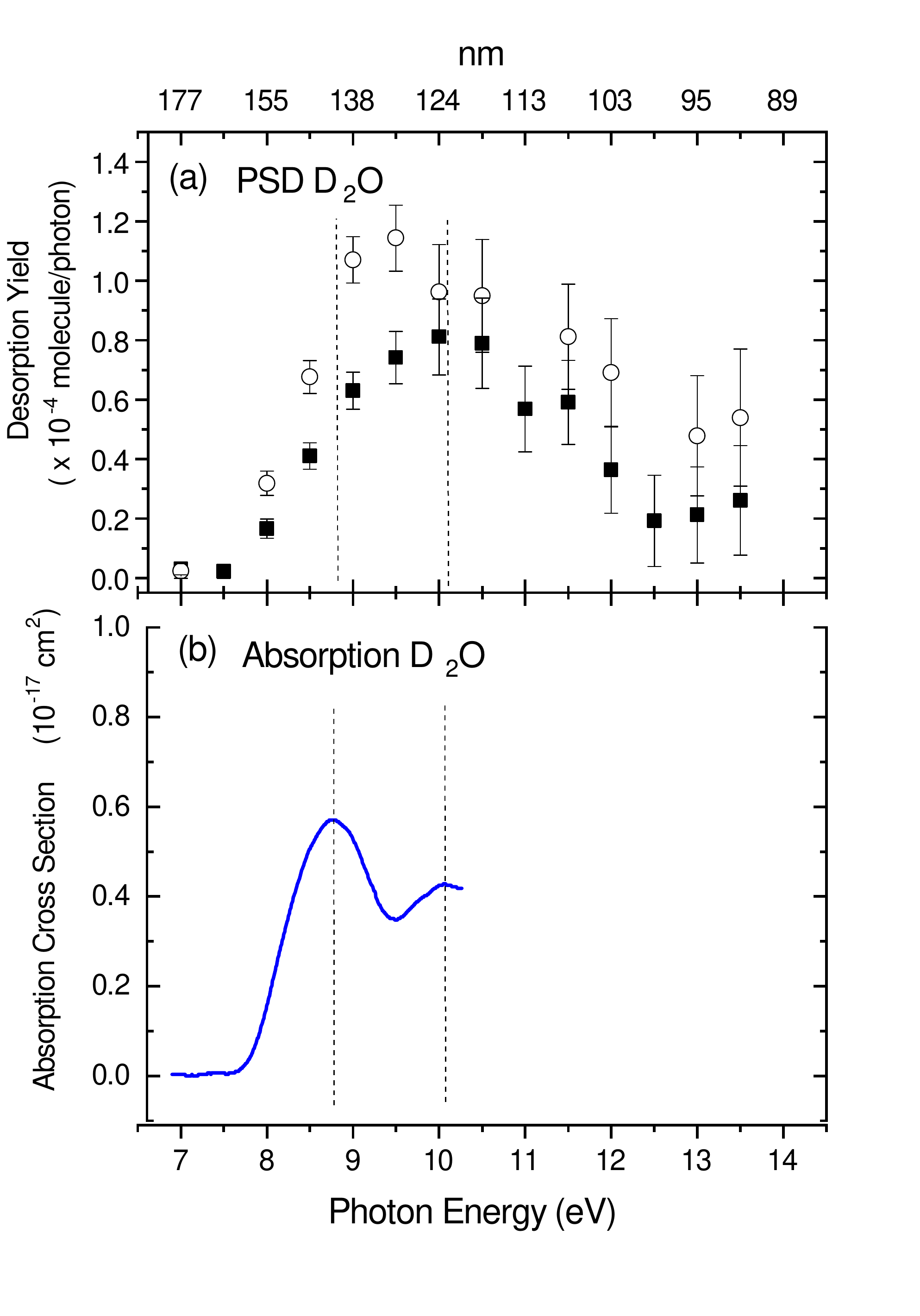}
   \caption{(a) Photodesorption yields (molecule per incident photon) of intact D$_2$O (black squares) and at 100 K (open circles) from 20 ML of c-ASW films (b) D$_2$O absorption cross sections from  \citet{cruz-diaz_vacuum-uv_2014}.The dotted vertical lines point the position of the first two excitonic bands.}
        \label{figD2O}
    \end{figure}

Figure \ref{figH2O}.a presents the energy dependence of H$_2$O photodesorption yields, hereafter called Photon-Stimulated Desorption spectra,
 from 7 to 13.5 eV obtained at 15 K and 100 K respectively. For comparison, Fig. \ref{figH2O}.b shows the absolute VUV absorption cross section of pure H$_2$O ice deposited at 8 K from 7.5 to 10.2 eV \citep{cruz-diaz_new_2018}. In order to get information at higher energies, part of the absorption spectrum calculated from the reflectivity spectrum of amorphous ice at 80 K \citep{kobayashi_optical_1983} has been rescaled on the same plot. In the gas phase, the absorption spectrum is characterised by two very broad absorption features centered around 7.4 eV and 9.7 eV  associated to the $3s4a_1 \leftarrow 1b_1 (\Tilde{A} ^1B_1 \leftarrow \Tilde{X} ^1A_1) $  and the $3s4a_1\leftarrow3a_1 (\Tilde{B} ^1A_1 \leftarrow \Tilde{X} ^1A_1)$ transitions respectively. They are the starting point of intense and sharper transitions toward series of bent/linear Rydberg states observed above 10 eV \citep{fillion_photodissociation_2001}. In the solid ice, taking a band picture view and comparing with the more  structured spectra of hexagonal ice, these two excited states can also be described as Frenkel-type excitons lying below the conduction band with transitions  at 8.5 eV and around 10.2 eV respectively (cf Fig. \ref{figH2O}), and with an associated ionization continuum starting around 11 eV \citep{kobayashi_optical_1983}.

The comparison of the figures shows that the H$_2$O PSD spectra do not simply mimic the absorption cross sections variations with photon energy. 
The PSD and absorption spectra present similar energy threshold (around 7 eV) but the first band peaking at 8.5 eV in the absorption spectrum is not
observed in the PSD spectra. This can be due to experimental spectral resolution limitation, because the bandwidth of the incident radiation and that 
of the absorption band are similar ($\sim$ 1 eV). The desorption yields are maxima around 10 eV, corresponding to the shoulder on the absorption spectrum assigned to the second exciton band \citep{kobayashi_optical_1983}. PSD and absorption spectra however, clearly deviate at higher energies (> 10 eV), for which the yields continue to decrease with photon energies despite significantly higher absorption cross sections above 10 eV (Fig. \ref{figH2O}a).  A similar 
behavior is seen for the desorption of D$_2$O (from D$_2$O c-ASW), as shown in Fig. \ref{figD2O}. In the first absorption band (8.75 eV), the 
desorption yields reach 70 \% of their maximum values.  The yields are maxima around 10 eV (second absorption band of the absorption spectrum).  Above 10 eV again, the desorption yields of D$_2$O are continuously decreasing with photon energies. 
 These effects are observed both at low (15 K) and high temperatures (100 K) for H$_2$O and D$_2$O (Fig. \ref{figH2O} and \ref{figD2O}). One can note this behavior is rather different from that observed from weakly bound condensates of diatomic molecules, such as CO, N$_2$ or NO ices, for which PSD and absorption spectra are extremely similar in spectral regions corresponding to electronic  transitions, as a signature of a Desorption Induced by Electronic Transition (DIET) process \citep{fayolle_co_2011, fayolle_wavelength-dependent_2013, dupuy_efficient_2017}.  In the present case, the deviation between PSD and absorption spectra likely results from multiple and/or less-direct desorption mechanisms at play associated with the dissociative character of the electronic excited states involved. The decrease of the photodesorption yields at high energies could result from the fact that the ionization of the sample does not contribute directly to the desorption events, whereas it contributes to the increase of the absorption cross section above the ionization threshold.
 
For both H$_2$O and D$_2$O, we observe no significant difference (< 5\%) in the intact water photodesorption yields recorded at the beginning of the measurement cycle of a PSD scan, compared to that obtained at the end.  This indicates the water photodesorption yields are not evolving with photon fluence ($\lesssim$ 6 $\times$10$^{16}$ photons.cm$^{-2}$). This could be surprising considering that VUV photons may induce structural and chemical modifications of the sample. Indeed, the porosity of amorphous water ice is known to decrease with photon fluence upon VUV (or ion) irradiation \citep{palumbo_h_2010}. In the present case however, since the ice has been grown at high temperature, the structure of the sample is already non-porous. Following \citet{Jenniskens_high-density_1995}, VUV irradiation merely favours the formation of high-density ASW, affecting the relative distribution of the coordination number of the hydrogen bonds. Possible slight modifications of the local density within the ice bulk do not seem to affect H$_2$O and D$_2$O photodesorption yields in the present experimental conditions.
 
 One major finding of the present investigation is to reveal that photodesorption is mostly efficient in the second excitonic band ( $\sim$10 eV), a result that 
 could not be anticipated  from the absorption cross sections and from our current understanding of the mechanisms, despite previous extended investigations on water. 
 Consequently, this result confirms, in retrospect, that excitation at Lyman-$\alpha$ (10.2 eV) is relevant to investigate water photodesorption in 
 the VUV. This is again in contrast with other systems, such as CO, N$_2$ or CO$_2$, that do not desorb efficiently at this energy and for which hydrogen-discharge lamps optimized for this transition are not very well-suited for a quantitative determination of the photodesorption yields in the VUV.

The temperature dependences of the photodesorption yields of intact water molecules from H$_2$O and D$_2$O ices are similar.
PSD spectra confirm that the desorption yields slightly increase with the sample 
 temperature. The integrated desorption yield from 7 to 13.5 eV increases by a factor $\sim$ 1.5 from 15 K to 100 K for both systems. This temperature dependence is weak as compared to what is observed for other molecules, as it will be shown below.
 
The general trends of the photodesorption yields with photon energy and temperature are 
very similar for both H$_2$O and D$_2$O, but their absolute values are dramatically different. 
Photodesorption yields of water measured at 10 eV (the maximum value over the total energy range) 
from 15 K c-ASW are (7$\pm$1) $\times$ 10$^{-4}$ molecule/photon and
 (8$\pm$1) $\times$ 10$^{-5}$ molecule/photon for H$_2$O  and D$_2$O  respectively.
 Considering the integrated desorption yields from 7 to 13,5 eV, 
 this effect is  characterized by the isotopic ratio Y(H$_2$O)/Y(D$_2$O)$\sim$ 10, independently 
 of the sample temperature.  We can therefore conclude that the desorption of H$_2$O is 
 significantly enhanced as compared to its heavier isotopologue D$_2$O,  at any photon energy 
 and temperature. This effect can not be explained by the difference in absorption cross section 
 between the two species since according to  \citet{cruz-diaz_vacuum-uv_2014}, H$_2$O absorption 
 cross sections exceed that of D$_2$O by $\sim$ 10\% only at 10 eV. It must originate mainly from different 
 weights of the various desorption channels at play producing intact water.

 \begin{figure}
     \centering
   \includegraphics [width=10cm]{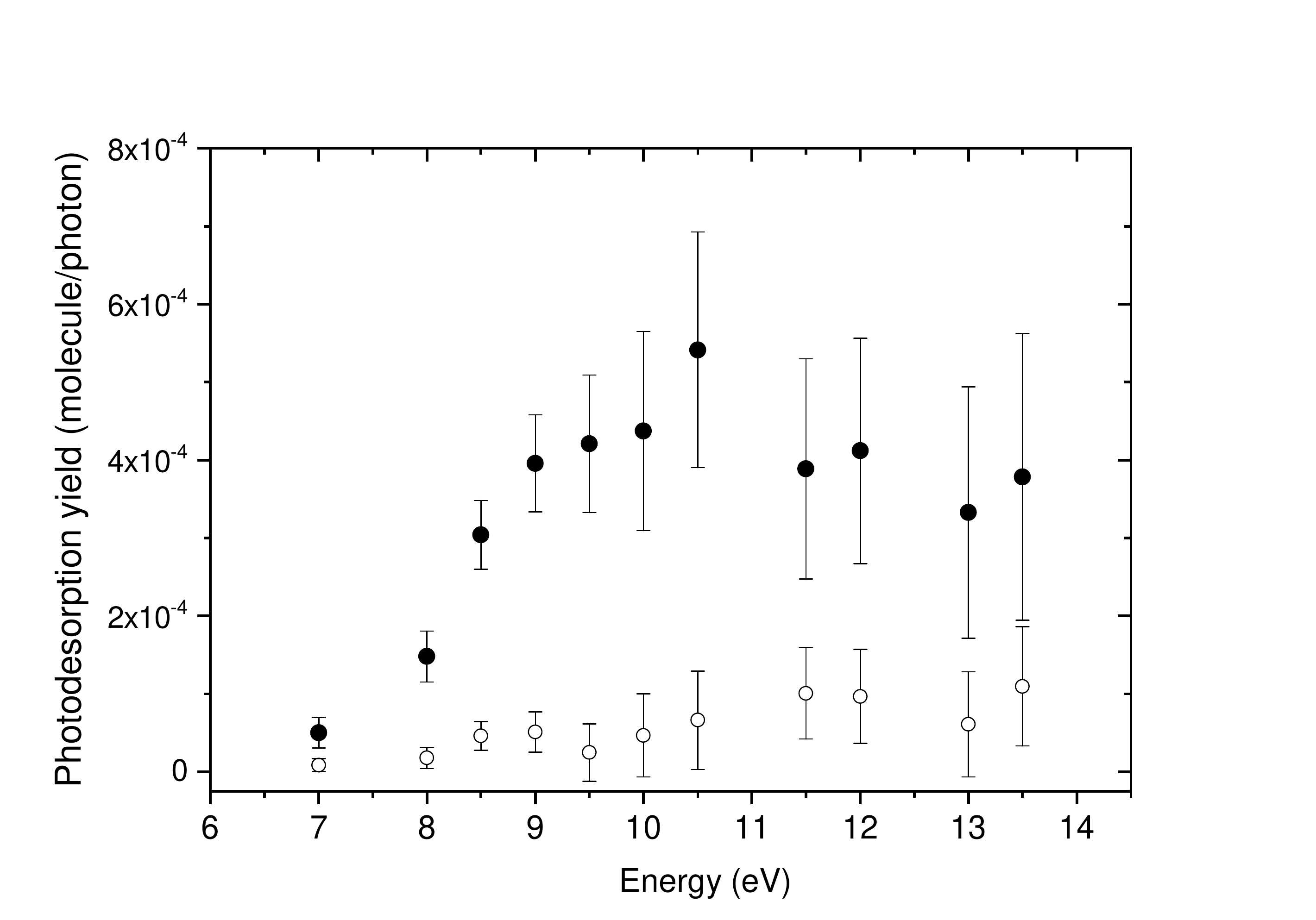}
   \caption{Photodesorption yields of OH (open circles) and OD (full circles) from 20 ML H$_2$O and D$_2$O compact amorphous ices at 15 K.}
        \label{figOH}
    \end{figure}

Figure \ref{figOH} presents the photodesorption of OH and OD radicals from H$_2$O and D$_2$O ices at 15 K respectively. 
Contrary to the previous case, no clear dependence of the photodesorption yields with ice temperature has been observed (cf Fig. \ref{figT}), and only
 low temperature results are shown in Fig.\ref{figOH}. No evolution of the photodesorption yields with photon fluence could be seen for OD nor for OH. The overall spectral shape of OD desorption seems different from that of D$_2$O. The OD desorption yields Y(OD) increase slowly with photon energy from 7 to 9 eV, reaching (3.9 $\pm$0.6) $\times$ 10$^{-4}$ molecule/photon at 9 eV. This desorption yield is maintained up to 13.5 eV. By contrast, the OH desorption yields Y(OH) are near the detection limit of this experiment, and if we consider the error bar, it is difficult to confirm its detection and draw any definitive conclusion about the evolution of OH desorption yields with photon energy. We estimate OH maxima desorption yields from H$_2$O are $\lesssim$ 10$^{-4}$ molecule/photon above 9 eV. 
By contrast to what was observed for the desorption of the parent water molecules, only the desorption yield of the heaviest radical (OD) is confidently measured, suggesting the existence of an isotopic-dependent mechanism.

\begin{figure}
     \centering
   \includegraphics [width=8.5cm]{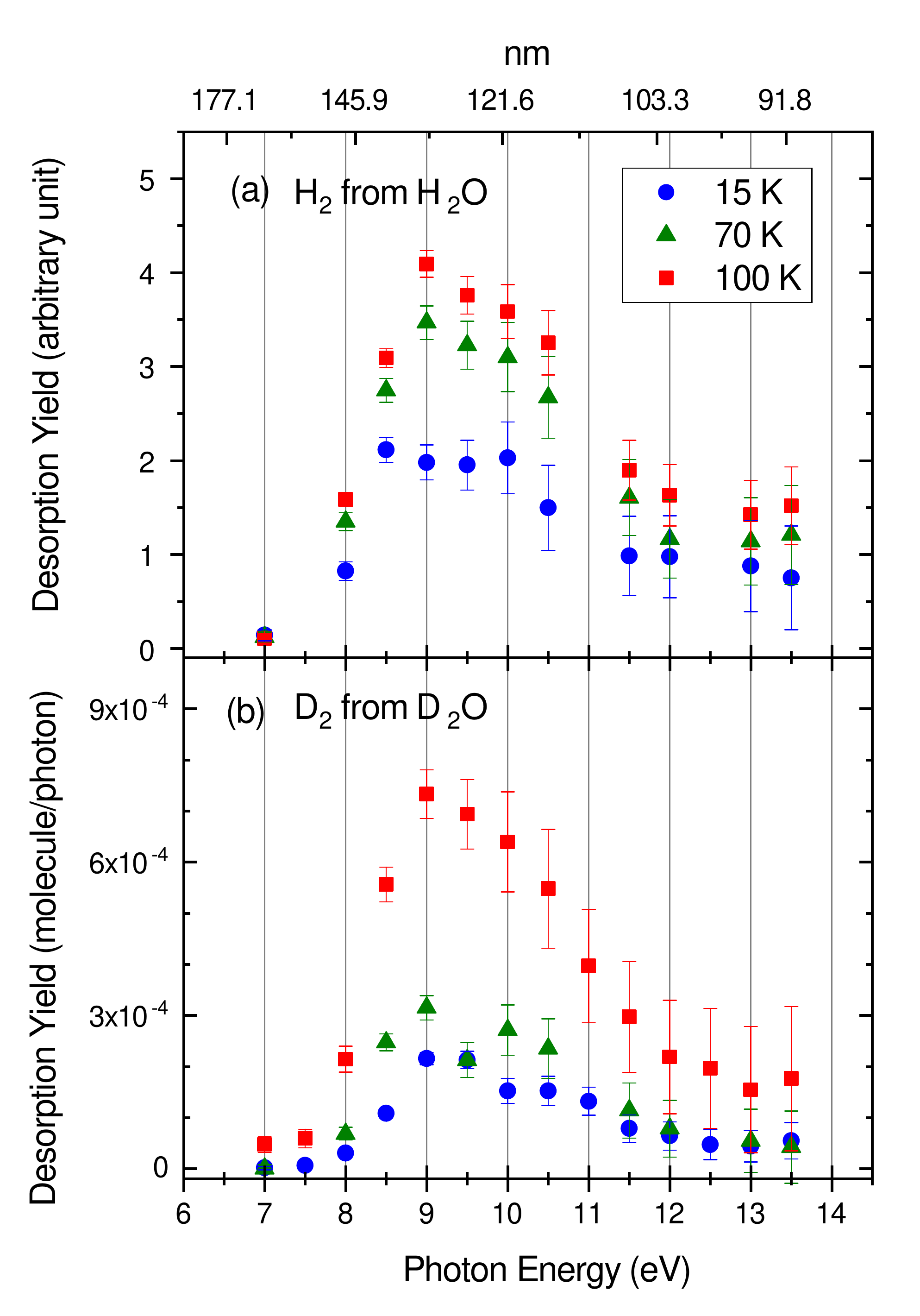}
   \caption{Photodesorption yields of molecular hydrogen at 15 K (blue circles), 70 K (green triangles), 100 K (red squares) for (a) H $_2$ from H$_2$O  (20 ML c-ASW) and (b) D$_2$ from D$_2$O  (20 ML c-ASW). The desorption yields for H$_2$ have been arbitrary scaled.}
        \label{figH2}
    \end{figure}

Figure \ref{figH2} shows the desorption of molecular hydrogen at several temperatures (H$_2$ and D$_2$ from H$_2$O and D$_2$O ices respectively). The shapes are similar to that of H$_2$O/D$_2$O desorption curves. All the desorption curves are peaking around 9 eV. This value almost matches the maximum of the first exciton band for both H$_2$O and D$_2$O.  The desorption curves measured at 30 K (not shown) were similar to that observed at 15 K, but a strong enhancement of the desorption yields is observed at higher temperature. From 15 K to 100 K, the desorption yields increase significantly, e.g. by a factor of 2 and more than 3 for H$_2$ and D$_2$ respectively. A more detailed investigation of the temperature dependence follows below.
 We observe a slow increase of H$_2$ and D$_2$ desorption signals with photon fluence (not shown), in line with substantial surface chemistry. From the D$_2$ signal, we estimate  this effect could account for a variation of 20\% of the yields measured over the timescale needed to record a spectrum (a variation not included in the error bars). In the present experiment, the mass 2 channel lies close to the mass cutoff of the QMS, which makes the apparatus function for this channel rather unreliable and prevents from any comparison between H$_2$ and D$_2$ photodesorption yields.

\begin{figure}
     \centering
   \includegraphics [width=8.5cm]{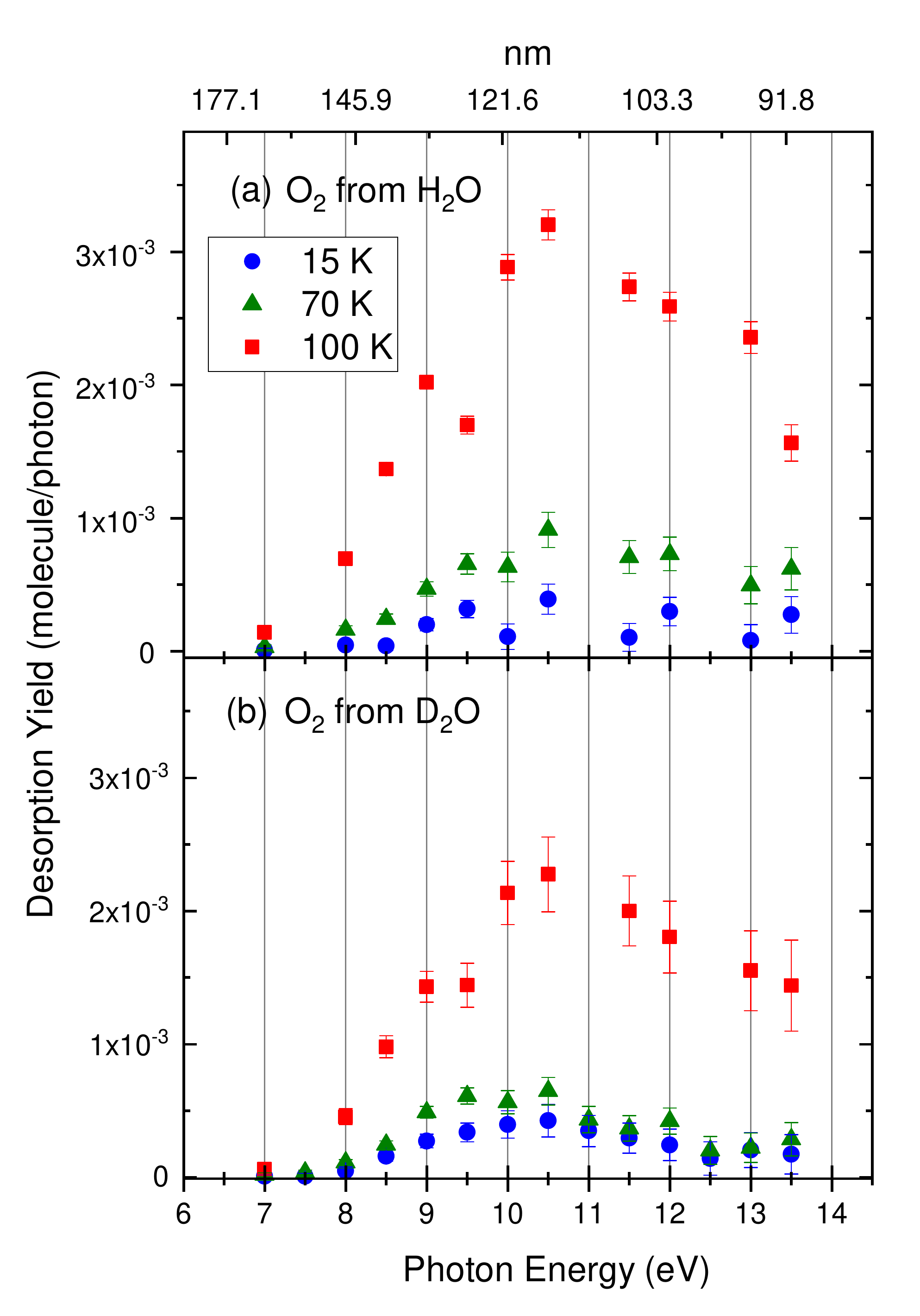}
   \caption{(Photodesorption yields of molecular oxygen at 15 K (blue circles), 70 K (green triangles), 100 K (red squares) for (a) O$_2$ from H$_2$O  (20 ML c-ASW) and (b) D$_2$ from D$_2$O  (20 ML c-ASW).}
        \label{figO2}
    \end{figure}

The desorption yields of O$_2$ from H$_2$O and D$_2$O ices are shown and compared in Figure 5a and 5b 
respectively. The PSD spectra present very similar trend, with a desorption increasing with photon energy and 
reaching its maximum around 10.5 eV. High desorption rates are maintained at higher energies, a behavior 
different from that observed for H$_2$ (D$_2$). The desorption of O$_2$ is known to be strongly dependent 
on the photon fluence \citep{cruz-diaz_new_2018}. We observed a non-linear increase over 
a long period of irradiation, without reaching any steady state after about 40 minutes of irradiation time 
($\sim$10$^{18}$ photons.cm$^{-2}$). 
Over the timescale needed to record a PSD spectrum, we observe that the O$_2$ production rates can increase
 by a factor of two. Consequently, the absolute desorption yield measured at one energy is sensitive to the number
of points that have been recorded previously to its determination (i.e. to the photon fluence). Although values obtained
 at each energy are resulting from an average of measurements taken in random order, care should be taken on the absolute 
desorption values obtained here for O$_2$.

Again, a significant enhancement of the desorption of O$_2$ with temperature is observed.  In order to get more
 insights about the role of this parameter, we investigated in more details
 the temperature dependence of the desorption yields for H$_2$O (D$_2$O), H$_2$ (D$_2$) and O$_2$ at 10 eV.
The results are shown in Figure \ref{figT}. The global behavior is surprisingly very similar for all the masses, with the 
notable exception of the radicals OD and OH. One can note two desorption components. A first component from 15 K to $\sim$ 60-80 K 
with nearly constant desorption yields. A second desorption component above $\sim$ 60-80 K, with desorption yields increasing linearly with T. 
Despite the fact that different mechanisms are underlying the desorption of these molecules, the behavior in the high temperature regime reveals a common source
for the enhancement of the various yields.

\section{Discussion}

In this experiment, the desorption of  H$_2$O (D$_2$O), OH (OD), H$_2$ (D$_2$) and O$_2$ are observed over 
the full energy range (7-13.5 eV). This observation is consistent with previous experimental photodesorption studies
involving specific wavelengths (121.6 nm, 157 nm or 193 nm). H is known to be by far the most abundant 
desorbing species from the ice \citep{andersson_photodesorption_2008} but the corresponding absolute yield can not be measured with our apparatus. 
In the present study, we focus on quantitative photodesorption yields of the main molecular compounds following the electronic 
excitation within the solid sample.  Among all the molecules detected, one can distinguish the desorption of water 
and hydroxyl radical presenting desorption channels that are partially connected to each other and for which advanced theoretical 
studies have been already proposed, to that of molecular hydrogen and molecular oxygen entirely associated to surface (photo)chemistry.
These two sets of molecules are discussed successively below. 

\begin{figure}
     \centering
   \includegraphics [width=8.5cm]{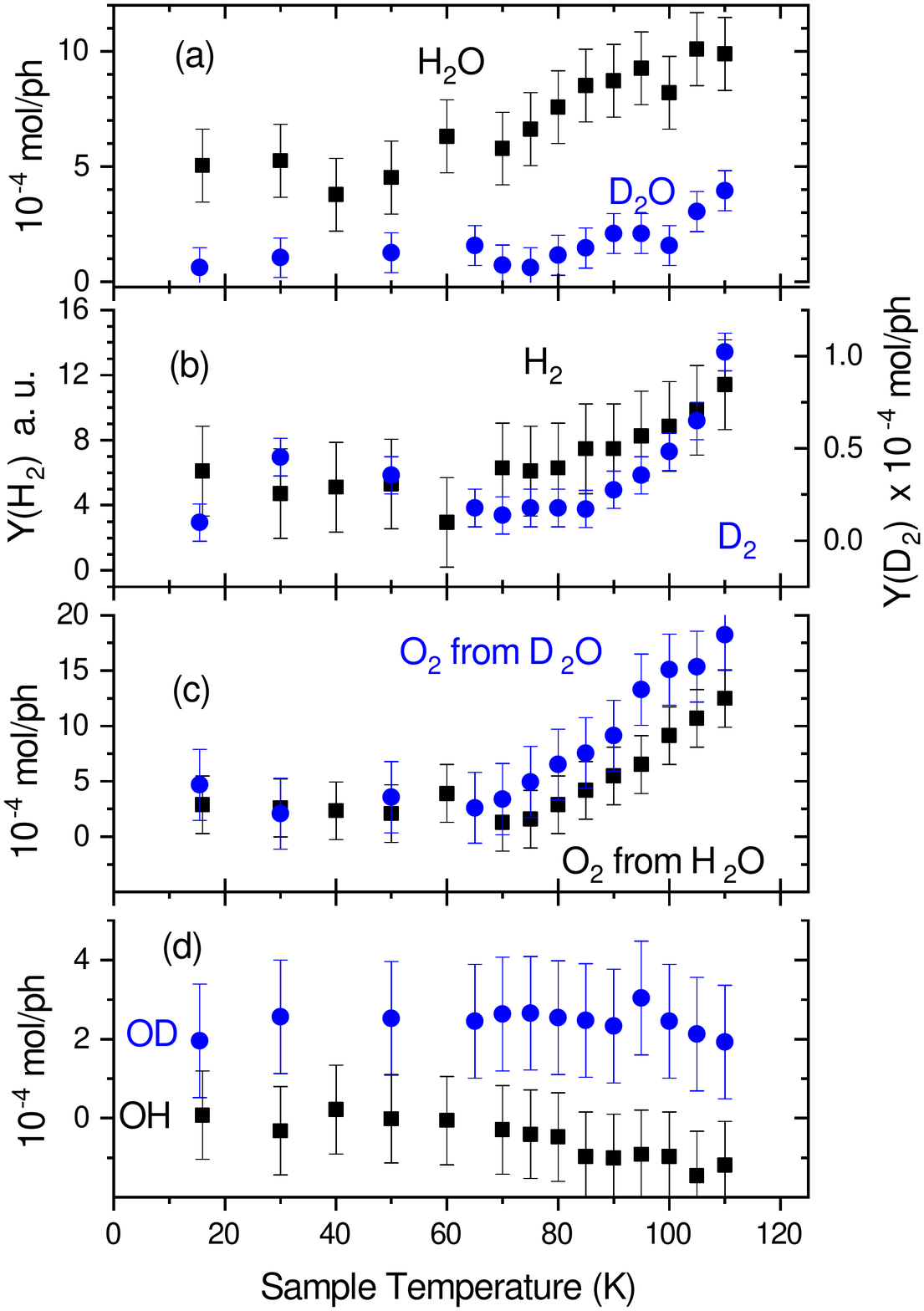}
   \caption{Photodesorption yields from H$_2$O (black squares) and from D$_2$O (blue circles) at 10 eV as 
                 a function of the sample temperature (100 ML c-ASW);  The desorbing species are : (a) water, (b) molecular hydrogen (c) molecular 
                 oxygen and (d) hydroxyl radicals.}
        \label{figT}
    \end{figure}

\subsection{Desorption of intact water molecules and hydroxyl radicals}

Several mechanisms have been proposed to explain the photodesorption of intact water : 
(1) (Electronic mechanism) excitons generated in the ice end-up localizing near the surface 
where the charge redistribution of the surface water molecules results in a repulsive electrostatic force
\citep{Nishi_photodetachment_1984, desimone_mechanisms_2013}.
(2) (Kick-out mechanism) a momentum transfer from an H atom released following water
 photodissociation to a surface water molecule which is ejected \citep{andersson_photodesorption_2008}
(3) (Chemical recombination) desorption after recombination of H and OH fragments originating from
 the same water molecule (geminate recombination) \citep{andersson_theoretical_2011}. Desorption
 after recombination at the surface of photoproducts from different water molecules (non-geminate 
 recombination) could also be considered in addition.

Our measurements do not present time dependency of the H$_2$O photodesorption yields, a fact that was also shown by \citet{oberg_photodesorption_2009} and attested by step-like behavior of the water desorption yields upon irradiation observed by \citet{cruz-diaz_new_2018}, indicating that if any photochemistry is at play, it must involve fast non-thermal reactions. Translational and rotational energy measurements of photodesorbed water molecules in their ground states from ASW at 90 K after irradiation at 157 nm were found in good agreement with those predicted by MD simulations for the kick-out mechanism at 10 K \citep{yabushita_translational_2009, hama_role_2010, andersson_theoretical_2011}. Another similar experiment at 108 K concluded that desorption on repulsive delocalized excitonic state was also likely to produce water in its ground state and could contribute to vibrationally excited water as well\citep{desimone_mechanisms_2013}. The recombination processes, for which no measurements of the translational and rotational energy were possible, are expected to produce vibrationally excited water molecules due to the high exothermicity of the reaction ($\sim$5 eV).
Thus, electrostatic repulsion, kick-out processes and surface chemistry are all believed to contribute to the photodesorption yields of intact water. \citet{andersson_photodesorption_2008} predict equivalent
efficiency for the kick-out mechanism and geminate recombination for the desorption of intact water. In our experiment, the relative contributions of each individual mechanism to the whole desorption process, including electronic processes and non-geminate recombination (or other exothermic chemistry) are not precisely known, but will be discussed in the following. 

The experiments exploring the desorption dynamics of OH radicals from ASW at 90 K following photon excitation at 7.89 eV (157 nm) have evidenced two different desorption mechanisms. The main one is characterized by non-thermal desorption (translational energy of 1300 K) due to direct water photolysis at the surface of ASW \citep{hama_formation_2009}.The second contribution arises from secondary photodissociation of H$_2$O$_2$ at high fluence (> 10$^{17}$ photons.cm$^{-2}$). This second mechanism is not dominant in the present study in which  the fluence is more limited. Therefore, the OH photodesorption signal results mainly from the direct photodissociation  of water, which is consistent with the absence of dependence with the surface temperature for this desorption channel (Fig. 6).
 
\begin{figure}
     \centering
   \includegraphics [width=8.5cm]{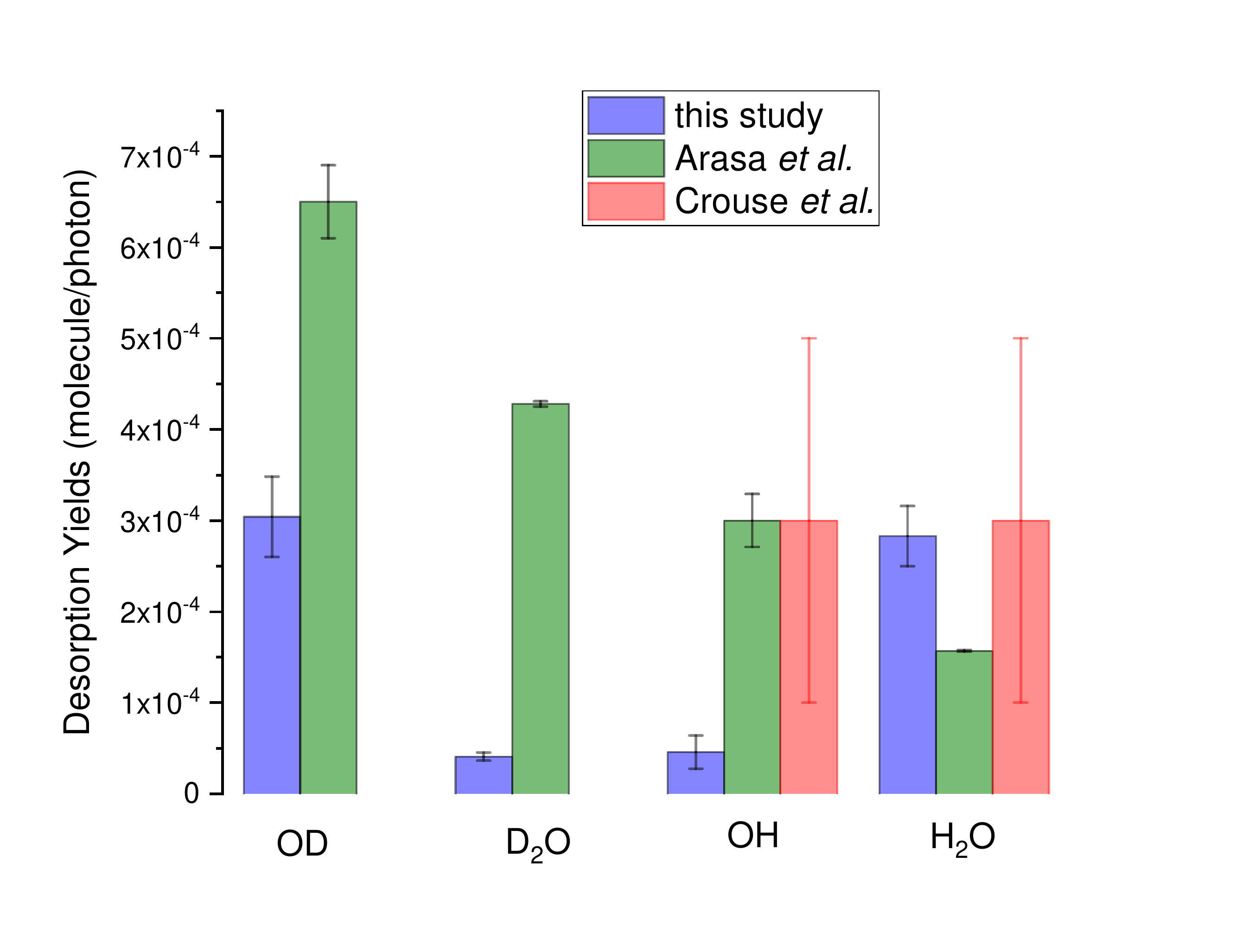}
   \caption{Experimental photodesorption yields for 8.5 eV photons energy at 15 K (blue) compared to molecular dynamics calculations from \citet{arasa_molecular_2011} at 10 K  (green) 
                and from \citet{crouse_photoexcitation_2015} at 11 K (red).}
        \label{figMD}
    \end{figure}
    
Molecular Dynamics (MD) calculations have investigated the photodissociation probabilities within the active top layers of the ice film. In order to compare the desorption yields  (molecule/photon) of the present study to these theoretical results, we have used the probabilities of desorption (molecule per monolayer)  $P^i_{des}$ given by \citet{arasa_photodesorption_2015} in each sublayer $i$.
The desorption yields are calculated by taking into account the probability $P^{ML}_{abs}$ that a photon is absorbed in a given ML and the probability that the photon has not been absorbed by the overlayers. For a given desorbing species, the total photodesorption yields,  following \citet{arasa_molecular_2011}, are obtained by summing over the top 4 layers since the contribution from deeper layers can be neglected :
\begin{equation}
  \centering
  y=\sum_{i=1}^4  P^i_{des}(1-P^{ML}_{abs})^{i-1}P^{ML}_{abs}
  \label{yield}
\end{equation}
 where $P^{ML}_{abs}=(1-e^{(-\sigma N)})$ is obtained using the absorption cross sections at 8 K measured  by \citet{cruz-diaz_vacuum-uv_2014-1} at the maximum of the first absorption band at 8.5 eV and taking N = 1.1 $\times$ 10$^{15}$ cm$^{-2}$ for ASW surface density \citep{kimmel_control_2001}. The comparison between theoretical yields from \citet{arasa_photodesorption_2015} and the measured desorption yields at 8.5 eV is presented in Figure \ref{figMD}. \citet{crouse_photoexcitation_2015} derived theoretical yields in a flexible water model allowing energy transfer between photofragments and the surrounding molecules. Their calculated values are shown for the non-deuterated species for comparison. One can note that calculated and experimental values lie remarkably within the same range. There are however several major differences between the experimental and theoretical values. At first, regarding the total yields Y$_D$=Y(OD)+Y(D$_2$O) and  Y$_H$=Y(OH)+Y(H$_2$O), we observe Y$_D$/Y$_H$=1 in the experiment and Y$_D$/Y$_H$=2 in the calculation. The isotopic effect observed on water desorption in the calculations originates from a most efficient momentum transfer for heavier species (kick-out mechanism).  Second, the distribution between OD (OH) and D$_2$O (H$_2$O) are different.  In the calculation, the desorption of hydroxyl radicals slightly dominates for both species with 60\% of OD and 40\% of D$_2$O on one side and 66 \% of OH and 34 \% of H$_2$O on the other side, or are predicted similar as calculated for OH and H$_2$O by \citet{crouse_photoexcitation_2015} (Fig. \ref{figMD}). 
Experimental data show an opposite behavior regarding the dominant desorption routes in the non-deuterated species, that is  88 \% of OD desorption (12 \% of D$_2$O) but $\gtrsim$ 86 \%  of H$_2$O desorption ($\lesssim$ 14 \% of OH). This results in a very large isotopic effect in the experimental yields characterized by Y(H$_2$O)/Y(D$_2$O)$\sim$10. The fact that Y(H$_2$O) > Y(OH) is consistent with  the data shown in \citet{cruz-diaz_new_2018}, although the absolute yields found in this latter study are higher than ours (an aspect which will be discussed below).\\

As mentioned earlier, the kick-out mechanism predicts an opposite behavior to that observed, with higher probabilities for the desorption of D$_2$O as compared to H$_2$O due to a better mass ratio between D and D$_2$O. This is a strong indication that other processes are active in the experiment.

Strong isotope effects observed in Electron Stimulated Desorption (ESD) from surfaces have been explained in the framework of Menzel-Gomer-Redhead (MGR) model \citep{Ramsier_electron-stimulated_1991}, which also applies to Photon Stimulated Desorption.  In this model, the excitation onto an isotope-independent excited potential energy surface is followed by fast deexcitation to highly vibrationally excited levels of the electronic ground state.
 The desorption occurs when the kinetic energy gained during this sequence is large enough to overcome the ground-state desorption barrier. The model predicts an isotope effect because a lighter particle is accelerated in a shorter time on the excited Potential Energy Surface (PES) and, thus, has less probability to relax onto the ground state before reaching the critical adsorbate-substrate distance on the excited-state above which the desorption becomes energetically allowed.  Quantitatively, for two species with masses m$_1$ and m$_2$, the isotope effect is given by
 \begin{equation}
 \centering
 $$\sigma_{m_1}/\sigma_{m_2}=(1/P_d)^{[\sqrt{m_2/m_1}-1]}$$ 
 \end{equation}
  where $\sigma_{m_1}$ and $\sigma_{m_2}$ are the respective desorption cross sections and P$_d$ the photodesorption probability of the mass m$_1$ species. \citet{wolf_dynamics_1991} have applied the MGR model to the isotopic fractionation of H$_2$O/D$_2$O on Pd(111) with $\sigma_{H_2O}/\sigma_{D_2O}\sim1.4$, a process mediated by the excitation of the metallic substrate which can not be transposable here. A variation of the MGR model has been proposed by \citet{zhu_vibration-mediated_1992} in order to explain the strong isotopic fractionation observed in the desorption of ammonia  ($\sigma_{NH_3}/\sigma_{ND_3}\sim 4$). In this model the considered repulsive PES is along the N-H coordinate instead of the adsorbate-substrate coordinate and a strong coupling between the N-H vibration mode to the N-surface mode is assumed to allow for the energy transfer leading to desorption.
  Excitation of water onto a repulsive state along the HO-H coordinate appears to be well-suited with such a picture due to the dissociative character of the water excited states. Given that m$_1$ and m$_2$ are the masses of D and H atoms in this revised model, $\sigma_{H_2O}/\sigma_{D_2O}$=10 gives a probability of desorption for H$_2$O P$_d$(H$_2$O)= 4$\times$10$^{-3}$ . 
  The corresponding desorption yield depends on the number of layers contributing to the desorption. Considering equation \ref{yield} to calculate the desorption yield and P$_d$(H$_2$O) for each layer, it ranges from 2.6$\times$10$^{-5}$ molecule/photon assuming a contribution of the 4 upper layers up to 5$\times$10$^{-4}$ molecule/photon in the case of 20 layers. A thickness of 11 ML would match the experimental yield of 2.8 $\times$10$^{-4}$ molecule/photon (obtained at 8.5 eV and 15 K). This value is realistic considering that electronic excitations might diffuse from the bulk to the interface \citep{petrik_electron-stimulated_2003}. Nevertheless, the validity of this modified MGR model in the case of water would need to be confirmed, because it requires a full energy transfer from the intra-molecular vibrational mode toward inter-molecular vibrational mode (that is the desorption coordinate). The efficiency of such a coupling between vibrational modes is questionable. In fact, strong dipole-dipole interactions between the OH groups results in a very fast vibrational energy transfer towards other intra-molecular vibrations of the water ice network \citep{zhang_ultrafast_2011} and we expect this effect to compete against desorption by favouring energy dilution into the matrix. Assigning such huge isotope effect to this mechanism alone seems thus doubtful, especially since it has not been backed up by previous characterisations of the desorption processes listed above, all contributing to the desorption yield \citep{andersson_theoretical_2011, hama_desorption_2010, desimone_mechanisms_2013}, but it cannot be completely ruled out. 
   
The role of chemical recombination can also be discussed despite the fact that no isotopic dependence was seen in the molecular dynamics simulations for this desorption channel \citep{koning_isotope_2013}. 
Indeed, it is tempting to correlate the differences observed in the desorption of hydroxyl radicals to that of H$_2$O and D$_2$O desorption. The desorption of OH/OD radicals originates from a direct photodissociation event, a fact further confirmed by the lack of temperature dependence in the OD signal (Fig. \ref{figT}).  As pointed out by \citet{koning_isotope_2013}, the probabilities of desorption of hydroxyl radicals (OH or OD) are mainly governed by their initial kinetic energy upon photodissociation. Indeed, considering an isolated water molecule, the conservation of energy and linear momentum implies that OD radicals have higher kinetic energies than the OH radicals upon water photodissociation. More precisely and following \citet{koning_isotope_2013}, for an initial photon energy of 8.5 eV, the total energy available after photodissociation is E=3.1 eV, which after distribution between X and OX (X= H or D)  photofragments and small corrections taking into account the average energy released in the vibration is partitioned as 0.275 eV and 0.142 eV of translational energy for OD and OH respectively.  This explains, in their simulations, higher yields for OD photodesorption with respect to OH and the independence with surface temperature. Whereas OD photodesorption is twice that of OH in the theoretical study, the differential desorption appears to be much more dramatic in the experimental data, in which OH signal is hardly decoupled from the noise level.
A recent molecular dynamics study of the energy dissipation on crystalline water ice surface which investigates the role of the initial kinetic energy on desorption probabilities clearly points out the presence of a threshold energy for desorption to occur \citep{fredon_energy_2017}, which is linked to the interaction energy between the desorbing species and the surface. The interaction energy of radicals in water clusters  (5 water molecules) were calculated around 0.26 eV, a value which is above the available kinetic energy for OH (0.142 eV)  \citep{allodi_hydroxyl_2006}.  Other quantum mechanics approaches (Own N-layer Integrated molecular Orbital Molecular Mechanic) applied to OH interacting on hexagonal water ice found a range of binding energies from 0.2 eV to 0.67 eV depending on the binding sites and the orientation of the OH radical. This set of binding energies are higher than the available kinetic energy for OH. A broad range of binding energies can be expected on compact amorphous water ice as well.  We suggest that the kinetic energy of most OH radicals is not sufficient to overcome their barrier, reducing drastically the  desorption probability.  Simultaneously, more OH photofragments, as compared to OD, remained at surface or in the ice and a part of them would be available for non-thermal surface reactions, thus resulting in H$_2$O/D$_2$O > 1, as observed. The OD desorption yields from D$_2$O irradiation at 8.5 eV is 3 $\times$ 10$^{-4}$ molecule/photon at 15 K a value similar to the H$_2$O yields (Fig. \ref{figMD}). This effect could thus account for a large part of the H$_2$O desorption, assuming most of the OH radicals are converted back to (desorbing) water molecules, and could contribute to the differences observed between H$_2$O and D$_2$O. \\

The desorption yields of water was previously estimated by different groups using microwave hydrogen discharge lamps.  \citet{westley_photodesorption_1995} reported a value of 3.5 $\times$ 10$^{-3}$ molecule/photon at 35 K and \citet{oberg_photodesorption_2009} derived a value of 2 $\times$ 10$^{-3}$ molecule/photon at 20 K. Assuming that most of the photon flux was at Ly-$\alpha$ emission, these values can be compared to the yields measured at 10 eV in the present study, that is (7$\pm$1)$\times$ 10$^{-4}$ molecule/photon and (8$\pm$1) $\times$ 10$^{-5}$ molecule/photon for H$_2$O and D$_2$O at 15 K respectively. As pointed-out by \citet{cruz-diaz_new_2018}, this apparent discrepancy is due to the experimental methods used in these previous studies (quartz crystal microbalance and infrared spectroscopy) that are both probing the photodesorption of water through its destruction, making difficult to separate the contribution of intact water desorption from other sources of water depletion from the solid phase (desorption as photofragments, photodissociation without desorption, chemical reaction at surface and within the bulk). \citet{desimone_mechanisms_2013} estimated a H$_2$O photodesorption yield of 1.8$\times$10$^{-4}$ molecule/photon from ASW at 108 K after laser irradiation at 7.9 eV (157 nm) and direct detection into the gas phase by time-of-flight spectroscopy, a value in very close agreement with this study which gives  (2.6$\pm$0.3)$\times$10$^{-4}$ molecule/photon at 8 eV.
 \citet{cruz-diaz_new_2018} recently reported new photodesorption yields after irradiation of ASW with a broad band microwave hydrogen plasma lamp and detection of water into the gas phase with a calibrated QMS, a detection method similar to the one used in the present study. Although a precise comparison would require to take into account the spectra profile of the lamp, the comparison reveals a gap between the two sets of data. At low temperatures (< 40 K) they differ by a factor $\sim$ 1.8 for H$_2$O, which is remarkably small, but by almost one order of magnitude for D$_2$O. The fact that the ratio are different between the two isotopologues suggests that the differences between the two groups can not simply be assigned to a shift in the absolute yields obtained after QMS calibration which would give a systematic gap  (although this could account for a factor of 2 typically). 
Also we point out the temperature dependence of the yields are different, since yields obtained in \citet{cruz-diaz_new_2018} increase significantly above 50 K, whereas that obtained in this work start increasing above $\sim$ 70 K. These two distinct thermal evolutions could be explained by two distinct sample morphologies. Indeed, the water ice samples are prepared at 100 K in the present study, producing compact and stable ASW, and they are prepared at 8 K in \citet{cruz-diaz_new_2018} and subsequently warmed up to the irradiation temperature, leading to highly porous ASW low density samples, with a morphology sensitive to the temperature due to pore collapse.  In porous-ASW, surface water molecules with low coordination number are more likely formed as compared to compact ASW and are distributed over a much larger surface area. These weakly bound species may also more easily desorb and contribute to higher desorption signal. It is generally assumed that compact forms of water are more relevant under interstellar conditions, in particular because of compaction induced by cosmic ion bombardment and VUV irradiation \citep{palumbo_formation_2006, mejia_compaction_2015, palumbo_h_2010}. Regardless, the differences observed between the two studies are still surprisingly high for D$_2$O and the role played by surface roughness is not easily predictable. This calls into question the role of the surface morphology in the relative contributions of the desorption mechanisms for water, and its consequence on fractionation. Indeed, molecular dynamics study have shown higher desorption probabilities of OH radicals from crystalline ice than from amorphous ice for example \citep{andersson_molecular-dynamics_2006}, a difference which may affect water desorption as discussed above.  It would be interesting to get a better estimate of the sensitivity of these desorption yields with the surface morphology in the experiments.

\subsection{Desorption of molecular hydrogen and molecular oxygen}

Although the formation and desorption of H$_2$ (D$_2$) and O$_2$ from the initial irradiation of water ice are not driven exactly by the same chemical pathways, they share common properties, resulting both from rapid diffusion of excitons to -or near- the surface  of ASW.  The production of molecular hydrogen from water ice can arise (a) directly from water photodissociation giving H$_2$ and O, or (b) indirectly by water photodissociation producing H and OH fragments followed by H atom recombination and/or H abstraction reactions with H$_2$O \citep{watanabe_measurements_2000}. In the gas phase, the first electronic excited state is purely dissociative.  In the condensed phase, H atoms can recombine with OH to form H$_2$O, limiting therefore the efficiency of water destruction within the bulk (cage effect), but not in the same way at the surface where the excited molecules are not entirely surrounded by the water matrix.  Molecular hydrogen released into the gas phase in highly excited states (v = 0-5; J = 0-17) has been detected after irradiation of 100 K ASW and polycrystalline ice, after excitation in the red-tail of the absorption band (193 nm, 157 nm) \citep{yabushita_release_2008, yabushita_measurements_2008}. The translational and rovibrational energy partition is consistent with the indirect mechanisms, with substantial internal excitation in the case of the recombination reaction, and low translational and rovibrational energy in the case of abstraction reaction.
The production of molecular hydrogen from amorphous solid water films on Pt(111) observed from electron-stimulated desorption (ESD) provides evidence for rapid migration of the electronic excitation deposited in the ASW film from the bulk to the ASW vacuum interface \citep{petrik_electron-stimulated_2003, petrik_electron-stimulated_2004}. The results thus suggest the molecules are produced at the interfaces of ASW and not from the interior of the film. 

All these processes occurring at the surface of ASW are initiated by the photodissociation of water.  The photodesorption yields of these fragments thus reflect the initial photodissociation efficiency. Assuming the photodesorption yields are proportional to the photodissociation efficiency, one can deduce from the shape of PSD spectra that most of the photodissociation occurs in the first absorption band, peaking around 9 eV, and that photodissociation cross sections are decreasing above 10 eV, despite the increase of the absorption cross sections at high energy (Fig. \ref{figH2}). 
 \\

Concerning molecular oxygen,  O$_2$ (X $^3\Sigma^-_g$) and O$_2$  (a $^1\Delta_g$) desorption have been observed from ASW at 90 K \citep{hama_role_2010} in the 157 nm laser irradiation experiment. The formation mechanism proposed involve O ($^3$P) atoms which react with OH to form O$_2$. O ($^3$P) atoms could arise from the recombination of two OH radicals or from the photodissociation of H$_2$O$_2$. Molecular oxygen originates from successive surface reactions where OH radicals are playing a central role. The ESD of molecular oxygen from amorphous solid water has also been subject of detailed investigations \citep{petrik_electron-stimulated_2006-1, petrik_electron-stimulated_2006}. The ESD experimental results could be reproduced with a kinetic model involving a multistep reactions sequence. In this scheme, OH radicals produced by the relaxation of the electronic excitation at the surface of ASW recombine to form H$_2$O$_2$,  which reacts itself with OH to produce HO$_2$.  HO$_2$ is believed to be the precursor of O$_2$ through a subsequent reaction with another species (most probably with OH arising from H$_2$O dissociation).
One can note the central role again played by mobile OH radicals at or near ASW vacuum interface, in all the above chemical schemes leading to O$_2$. The fact that we observe a strong dependence of O$_2$ photodesorption yields with irradiation time is consistent with the accumulation of OH and H$_2$O$_2$ photoproducts on the surface and with these types of secondary chemistry. It is also true for our measured temperature dependence of O$_2$ desorption (see the next section).

\subsection{Temperature effects}

Both molecular hydrogen and molecular oxygen are secondary processes initiated by successive photodissociation events. At low temperature and following the pictures given in the previous section, these molecules are believed to be formed near or at the  ASW/vacuum interface. The desorption of these two molecules is significantly enhanced at high temperatures. This increase can not be assigned to enhanced surface diffusion. Indeed, at high temperature, the residence time of the particles are greatly reduced, limiting therefore the probability of reaction. This competition between diffusion and desorption is known for example to explain why the formation of molecular hydrogen via the Langmuir-Hinshelwood mechanism is efficient in a tiny range of temperatures \citep{vidali_h2_2013}. Consequently, higher photodesorption yields are not explained by phenomena occurring entirely at the external surface of the sample, but must involve particles originating underneath. The fact that all the desorbing species present a threshold in the same temperature range around $\sim$ 60-80 K, as shown in Fig. \ref{figT} suggests an origin linked to the water matrix itself. Indeed, the diffusion of species within the water ice network has been shown to be similar to self-water diffusion inside the matrix, due to the fact that diffusion is governed by ice restructuration \citep{ghesquiere_reactivity_2018}. ASW crystallizes to cubic ice above 125 K. Below 100 K, ASW is continuously evolving. The morphology of porous ASW samples investigated by infrared spectroscopy which probes the dangling OH bonds at the surface of micropores have shown evidence for ice restructuration near 75-80 K \citep{rowland_probing_1991}.  Although the experimental protocol used in the present experiment leads to the production of compact structures that are expected to be nonporous, \citep{stevenson_controlling_1999}, \citet{rowland_probing_1991} study suggests that the ice matrix gets sufficient energy to transform around 80 K. This would allow the atoms and molecules trapped underneath the surface to gain mobility, react and/or reach the external surface of the sample. One can note in addition that thermal desorption of H$_2$ and O$_2$ occurs around 10-20 K and $\sim$30 K respectively. Any of these molecules reaching the surface above 70 K would thus promptly desorb without the need of non-thermal energy. By contrast, thermal desorption of water isn’t efficient even at 110 K and exothermic events near the surface of the sample are still required to account for intact water desorption. In other words, the stronger enhancement of O$_2$ and H$_2$ (D$_2$) as compared to water, can be explained by a desorption thermally supported for the most volatiles species. 

\subsection{Astrophysical implications}

 The HDO/H$_2$O and D$_2$O/HDO abundance ratios derived from observations in star forming regions are good tracers of water formation. Indeed, these ratios are used to follow the water trail all along the way from the prestellar stage up to the formation of disks, planetesimals and planets, and provide valuable keys to try to understand to what extent molecules formed in interstellar clouds can be preserved \citep{van_dishoeck_water_2014, ceccarelli_deuterium_2014}. The D$_2$O/HDO ratio, in particular, has been shown to be a a pertinent probe of the origin of water into protostellar disks \citep{furuya_water_2017}, giving some indication of a presolar origin of water in 67P comet \citep{altwegg_cometary_2019}. Quantitative predictions provided by astrochemical models are thus crucial to interpret the isotopic abundances. In this context  it is important to provide good laboratory data and to understand how the photodesorption could affect  fractionation.
 
 In this experimental study, we have explored the absolute photodesorption yields of water on a wide energy range relevant to star-forming regions. This provides the first complete picture of the yields variation in the VUV. Concerning D$_2$O,  photodesorption yields from D$_2$O diluted into H$_2$O-rich ice would have been required to be fully relevant to interstellar conditions. However \citet{koning_isotope_2013} have shown that the desorption probabilities of D$_2$O embedded in H$_2$O ice are in fact very similar to that of D$_2$O in D$_2$O ice at any temperature. This indicates that the effect of the kick-out and recombination mechanisms are similar for both systems. Although, it is not known how the potential electronic relaxation (MGR model) contribution would change in the case of D$_2$O molecules embedded into an H$_2$O matrix, we believe that our measurements provide valuable estimates of D$_2$O photodesorption yields.  One should also keep in mind the high range of values associated to these measurements of photodesorption yields, as attested by the deviations observed between different experimental studies (up to one order of magnitude), as mentioned in the previous section. 
 
 In order to take into account the variation of the photodesorption efficiency with photon energy  over the 7-13.5 eV energy range, new photodesorption yields have been calculated for both systems by taking into account the spectral profiles of the Interstellar Standard Radiation Field (ISRF) and that of secondary emission induced by cosmic rays. The results are shown in table \ref{table:1}, which includes in addition the yields for an excitation at 10 eV ($\sim$ Ly-$\alpha$) for comparison. The sensitivity of the yields to different spectral profiles is limited because the maximum of photodesorption lies at Ly-$\alpha$ energy corresponding to the main emission induced by cosmic rays and because the ISRF profile is relatively flat in this energy range.
 We recommend to use in astrochemical models  $5 \times 10^{-4}$ molecule/photon at low temperatures (< 60 K) for intact H$_2$O. More precisely, we found that the temperature dependence above 60 K can be described by a simple linear function of the temperature T (in Kelvin). The photodesorption yields for H$_2$O are :
\begin{align}
  Y(T) &=5 \times 10^{-4} \textrm{ molecule/photon   	 (T} \leq \textrm{60 K)} \nonumber \\
  Y(T) &=5 \times 10^{-4} + 1.1 \times 10^{-5}\times(T-60)  \textrm{ molecule/photon}  \nonumber \\
  &\textrm{(60 K < T} \leq \textrm{ 110 K)} 
  \end{align}

  The corresponding D$_2$O photodesorption yields can be described by the same law but must be divided by 10.
 
 The determination of isotopic abundances in the solid phase of water in observations is very difficult due to the low-level of absorption signals from amorphous ice mantles \citep{dartois_revisiting_2003} and the presence of other solid-phase interstellar molecule vibrational features close to the O–D stretching mode band frequency \citep{urso_solid_2018}. Besides, the observations of gas-phase D$_2$O remain very scarce and difficult \citep{coutens_high_2014}.  It is therefore important to know whether the abundances ratio measured in the gas phase would reflect that in the solid phase. The present results show that in cold regions where the UV photodesorption dominates the gas phase water abundances, the H$_2$O/ D$_2$O abundances ratio measured in the gas phase is not reflecting directly that of the solid phase value because the photodesorption yields largely differ. 
 
 %
%
\begin{table}
\caption{VUV photodesorption yields ($\times$ 10$^{-4}$ molecule/photon) of water from c-ASW. $^a$from \citet{mathis_interstellar_1983} ; $^b$from \citet{gredel_cco_1987}}           
\label{table:1}      
\centering                          
\begin{tabular}{c c c c c}        
\hline\hline                 
\\
   & 15 K  & 100 K & 15 K  &  100 K \\
 Energy (eV) & H$_2$O & H$_2$O & D$_2$O &  D$_2$O  \\    
 
\hline                        

\\
 ISRF$^a$ &  $4.0\pm0.7$  &  $6.1\pm0.7$  &  $0.4\pm0.1$  &  $0.7\pm0.1$  \\      
Second. UV $^b$ &  $5.5\pm0.9 $  & $7.6\pm0.9 $ &  $0.6\pm0.1 $ &  $0.9\pm0.1 $\\ 
 10.2 eV (Ly-$\alpha$)  & $6.7\pm1.3 $  & $8.8\pm1.0 $ &  $0.8\pm0.1 $ &  $1.0\pm0.2 $\\ 
    \\
\hline                                   
  
\end{tabular}
\end{table}

 The experimental study of HDO photodesorption is very challenging and has not been addressed here. Regarding the electronic relaxation model (MGR), the desorption probability of HDO would give lower desorption when the O-D stretch mode is initially excited as compared to O-H,  which statistically results in a lower desorption probability of HDO as compared to H$_2$O. Considering the chemical recombination route, this study shows that depending on their kinetic energy, the hydroxyl radicals OX (X=H or D) promptly desorb following water photodissociation or remain in the solid and lead to X$_2$O desorption. 
 In HDO, by contrast with the relative behavior between H$_2$O and D$_2$O photodissociation, it is the OH fragments which gain more kinetic energy upon HDO photodissociation as compared to OD photofragments, which should result (by extrapolation of our result) to a differential photodesorption of OH as compared to OD radicals. This effect is modulated by the HDO photodissociation branching ratio, which gives two (H+OD) fragments for one (D+OH) fragments \citep{koning_isotope_2013}. 
Thus, regarding  the chemical recombination route, we predict the photodesorption of HDO to be 2/3 that of H$_2$O. Both channels (electronic relaxation and chemical recombination) are thus contributing to deuterium fractionation, although these effects remain small  considering the relatively low fractionation in water ices derived from IR observations (HDO/H$_2$O $\lesssim$ 10$^{-2}$)  \citep{parise_search_2003, dartois_revisiting_2003}.  
   Recent progresses in physical and astrochemical models have provided consistent scenarios explaining the evolution of the HDO/H$_2$O and D$_2$O/HDO ratio at the early stage of molecular clouds, in protostellar cores \citep{taquet_multilayer_2014, furuya_water_2015, furuya_reconstructing_2016, kalvans_chemical_2017} and during the incorporation of water into disks \citep{furuya_water_2017}.  As initially proposed by \citet{dartois_revisiting_2003} and further demonstrated in recent models, H$_2$O is formed primarily at the early stages of molecular clouds, whereas HDO and D$_2$O are mostly formed later in the very cold and high density stage, where UV irradiation is less effective and the top layers of ice mantle enriched with many other species (CO, CO$_2$, CH$_3$OH, NH$_3$ etc..). 
      Thus, electronic relation and/or chemical recombination could impact the ice mantle composition during its growth, favoring the fractionation in its bulk.
   Regarding the highly deuterated top layers of the mantle, as our results suggest that the photodesorption of water is critically sensitive to the surface morphology and composition,  the role played by water photodesorption from such complex chemical environment still needs to be studied to better constrain its effect on fractionation.

In this study, we found that at least  86\% of H$_2$O desorbs as intact water and not as OH+H photofragments : contrary to what is generally assumed, we found the OH photodesorption yield is weak. This finding is paramount for the chemistry in the comet-formation region of protoplanetary disks. Indeed, using time-dependent gas-grain chemical model in a region of a T-Tauri disk corresponding likely to comet-formation, \citet{chaparro_molano_role_2012} have shown that  secondary VUV (induced by cosmic rays) photodesorption from water ice delivers highly reactive OH radicals into the gas phase. This pathway opens new gas phase chemical routes that change the chemical balance and the final abundances, enabling in particular oxygen to be distributed in O$_2$, SiO and atomic oxygen, and carbon to be stored in CO and CO$_2$. 
As our results show that much less OH is produced by photodesorption than previously calculated, the model predicts that in poor-OH environment, most of oxygen will be found in H$_2$O, CO and CO$_2$ and a fraction of carbon in CH$_4$, which matches much better the cometary composition. Also it is important to note that the production OH radicals into the gas phase arising from water ice photodesorption does not reflect that of direct gas phase photodissociation, which can give large errors in the estimates of photodesorption yields derived from gas phase photodissociation cross sections data \citep{fogel_chemistry_2011}.

\section{Conclusion}

The photodesorption yields of H$_2$O, OH, H$_2$ and O$_2$ from amorphous H$_2$O ices, D$_2$O, OD, D$_2$ and O$_2$ from amorphous D$_2$O ices, have been measured 
from 7 to 13.5 eV and for sample temperatures between 15-110 K. This provides the first photon-energy dependence of the photodesorption yields in the vacuum-UV.
 The yields were found to increase rapidly from 7 to  10 eV and to decrease at higher energies. For intact water desorption, the maximum desorption yields are found in the second exciton band
 around 10 eV.
 
 The desorption at low (15 K) and high (100 K) temperatures present very similar photon energy dependence for all species, with the exception of hydroxyl radicals.  For the desorption of  H$_2$O (D$_2$O), and to a
greater extent that of H$_2$ (D$_2$) and O$_2$, we observe a dramatic enhancement above (70$\pm$10) K. This effect is assigned to the contribution from material originating deeper underneath the surface 
contributing to the chemical processes at higher temperatures including desorption induced by exothermic reactions for H$_2$O and D$_2$O, along with desorption thermally assisted for the lighter species.
 
The recommended H$_2$O photodesorption yields over the VUV range are (5$\pm$2)$\times$10$^{-4}$ molecule/photon
from compact ASW at low temperature (15 K) and increase linearly above 60 K with a rate of 1.1$\times$10$^{-5}$ molecule/photon/K.
 The corresponding D$_2$O photodesorption yields were found to be ten times lower. The relative desorption of OH/OD photofragments, associated to direct photodissociation are anti-correlated to that of H$_2$O/D$_2$O.  These effects can be explained by a differential chemical recombination involving OH (OD)  photofragments and/or possibly by a Menzel-Gomer-Redhead isotope effect in desorption. In the former case, we suggest that the trapping of OH photofragments on our c-ASW sample is converted into H$_2$O desorption, a process making the relative weights of the photodesorption
channels sensitive to the isotopic composition and surface morphology. This process could contribute to increase the HDO/H$_2$O ratios
in the solid phase during the build-up of the water ice mantles in molecular clouds and to favor adequate chemical environment
 for oxygen bearing molecules in comet-forming regions of protoplanetary disks.

\begin{acknowledgements}
    We acknowledge SOLEIL for provision of synchrotron radiation facilities under the projects 20150760 and 20180060. We thank Laurent Nahon and the DESIRS team for their assistance on the beamline. This study was 
    supported by the Programme National Physique et Chimie du Milieu Interstellaire (PCMI) 
    of CNRS/INSU with INC/INP co-funded by CEA and CNES. It was done in collaboration and with 
    financial support by the European Organization for Nuclear Research (CERN) under the collaboration 
    agreement KE3324/TE.  Financial support from the LabEx MiChem, part of the French state
    funds managed by the ANR within the investissements d'avenir program under reference ANR-11-10EX-0004-02,
    and by the Ile-de-France region DIM ACAV program, is gratefully acknowledged.
\end{acknowledgements}

\end{document}